\documentclass[12pt]{iopart}
\usepackage{iopams}
\usepackage{setstack}
\usepackage{amsfonts}
\usepackage{float}
\usepackage{placeins}
\usepackage{cite}
\usepackage{graphicx, graphics, epsf}
\usepackage{xcolor}
\usepackage[latin1]{inputenc}

\begin{document}

\jl{1}%

\title{Elastodynamics and resonances in elliptical geometry}%

\author{S. Ancey, E. Bazzali\footnote[1]{ebazzali@univ-corse.fr}, P. Gabrielli\footnote[2]{gabrieli@univ-corse.fr} and M. Mercier}%

\address{UMR CNRS 6134 SPE, Universit\'{e} de Corse, Facult\'{e} des
Sciences, 20250 CORTE, FRANCE}%

\begin{abstract}
The resonant modes of two-dimensional elastic elliptical objects are studied
from a modal formalism by emphasizing the role of the symmetries of the
objects. More precisely, as the symmetry is broken in the transition from
the circular disc to the elliptical one, splitting up of resonances and
level crossings are observed. From the mathematical point of view, this observation can be explained by
the broken invariance of the continuous symmetry group ${\mathcal{O}(2)}$ 
associated to the circular disc. The elliptical disc is however invariant
under the finite group ${\mathcal{C}}_{2v}$ and the resonances are
classified and associated with a given irreducible representation of this
group. The main difficulty stands in the application of the group theory in
elastodynamics where the vectorial formalism is used to express the physical
quantities (elastic displacement and stress) involved in the boundary
conditions. However, this method significantly simplifies the numerical treatment of the problem
which is uncoupled over the four irreducible representations of ${\mathcal{C} }_{2v}$. This provides a full classification of the resonances. They are
tagged and tracked as the eccentricity of the elliptical disc increases.
Then, the splitting up of resonances, which occurs in the transition from
the circular disc to the elliptic one, is emphasized. The computation of
vibrational normal modes displacement also highlights mode splittings. A
physical interpretation of resonances in terms of geometrical paths is
provided.
\end{abstract}%



\maketitle%

\section{Introduction}

\label{intro}

Resonant modes of elastic objects have been extensively studied in the last
decades, more particularly for simple shapes \cite{love}. In 1882, Lamb was
the first to solve exactly the problem of the free vibrations of an elastic
isotropic and homogeneous sphere \cite{lamb}. Visscher \textit{et al.} \cite%
{visscher} introduced a method based on Hamilton's principle approach to
compute resonant modes of elastic objects for various geometries. More
recently, Saviot and Murray \cite{saviot} focused on the classification of
the spheroidal vibrational modes of an elastic sphere. In the context of
classical ray dynamics, Tanner and Søndergaard studied the eigenfrequencies
of an elastic circular disc \cite{sondergaard} and their connection to the
periodic rays in the circular domain. Little attention has been given to the
elliptical geometry in elastodynamics comparatively to acoustic scattering 
\cite{bowman}. However, many techniques ($S$ matrix \cite{VaradanPao}, modal
formalism \cite{leon}) employed in this latter context can be transposed to
the interior problem.

In this paper, the elastodynamic resonances of elliptical objects are
studied from a modal formalism. In the scalar case, by using the appropriate
elliptic coordinates, the Helmholtz equation separates into two equations
involving Mathieu functions \cite{mechel , stephane}. Unfortunately, in
elastodynamics, the Helmholtz equation does not separate in elliptic
coordinates due to the existence of longitudinal and transverse waves \cite%
{VaradanPao, Varadan}. This brings us to use circular cylinder coordinates $%
(\rho ,\theta ,z)$. Then we shall assume that the problem is independent of
the $z $-coordinate and thus reduces to a two-dimensional one, referred as 
\textit{plain strain}, the elliptical disc, described by the polar
coordinates $(\rho ,\theta )$.

Several authors have paid attention to the role of the symmetries in various
contexts \cite{wirzba, yves ,paul, stephane}. The present work deals with
the resonant modes of the elastic elliptical disc, focusing on the splitting
up of resonances which occurs in the transition from the circular disc to
the elliptical one. In terms of group theory, this corresponds to the
symmetry breaking $\mathcal{O}(2)$ $\rightarrow $ $\mathcal{C}_{2v}$. This
splitting up has been numerically observed for the first time by Moser and Ü%
berall \cite{moser}. More recently, Chinnery and Humphrey discussed about
mode splittings and level crossings in the study of the acoustic resonances
of a submerged fluid-filled cylindrical shell \cite{chinnery}. However, none
of these authors provides an explanation or analytical description to this
phenomenon. In the scalar case, Ancey \textit{et al} \cite{stephane,theseste}
have highlighted and explained the splitting up of resonances in the
elliptical geometry using a method involving group theory \cite{morton}.
This technique has also been used to study multiple scattering \cite%
{yves,paul}. The main advantages of this method stand in (i) the uncoupling
of the equations, (ii) the classification of resonances, (iii) the
highlighting of the splitting up of resonances and its interpretation in
terms of symmetry breaking.

The paper is organized as follows. In section 2, the geometry of the problem
is presented and algebraic considerations are recalled. Then, from the
boundary condition, four systems of equations are obtained, each one
associated with a given irreducible representation of the symmetry group $%
\mathcal{C}_{2v}$. The problem is uncoupled and these systems can then be
solved numerically by truncation and used to obtain the resonances. In
section 3, numerical results are presented for the resonances. They are
tracked as the eccentricity of the elliptical disc increases and splitting
up of resonances is emphasized. The computation of vibrational normal modes
displacement also highlights the mode splittings. Finally, a physical
interpretation in terms of periodic orbits is given and the phenomenon of
mode conversion is observed.

\section{Mathematical Formalism}

\subsection{Position of the problem}

Let us consider an elastic elliptical disc. The elastic medium is
characterized by the longitudinal and transverse velocities $c_{L}$ and $%
c_{T}$ and we introduce the wave numbers $k_{L}=\omega /c_{L}$ and $%
k_{T}=\omega /c_{T}$, where $\omega $ is the angular frequency. The time
dependence $e^{-i\omega t}$ is assumed throughout the paper. As explained in
section \ref{intro}, we shall use the polar coordinates ($\rho $, $\theta $%
). The geometry as well as the notations used are displayed in figure \ref%
{fig:ellipse}. The elliptical boundary is a closed curved described by the
radius $r(\theta )$ with a continuously turning outward normal $\mathbf{n}%
(\theta )$ defined by%
\begin{equation}
\mathbf{n}(\theta )=n_{\rho }\mathbf{e}_{\rho }+n_{\theta }\mathbf{e}%
_{\theta }  \label{normale}
\end{equation}%
with 
\begin{equation}
n_{\rho }=\frac{1-e^{2}\cos ^{2}\theta }{\sqrt{1+e^{2}(e^{2}-2)\cos
^{2}\theta }},\;\;\;\;\;n_{\theta }=\frac{e^{2}}{2}\frac{\sin 2\theta }{%
\sqrt{1+e^{2}(e^{2}-2)\cos ^{2}\theta }}
\end{equation}%
and 
\begin{equation}
r(\theta )=\frac{b}{\sqrt{1-e^{2}\cos ^{2}\theta }}
\end{equation}%
where $e^{2}=1-\frac{b^{2}}{a^{2}}$ defines the eccentricity $e$ (figure \ref%
{fig:ellipse}). 
\begin{figure}[h]
\includegraphics[scale=0.7]{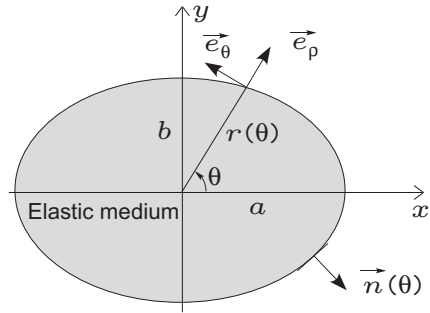}  \centering
\caption{Geometry of the problem.}
\label{fig:ellipse}
\end{figure}

The elastic displacement $\mathbf{u}$ is expressed using the Helmholtz
decomposition: 
\begin{equation}
\mathbf{u}=-\mathbf{\nabla }\phi +\mathbf{\nabla }\wedge \mathbf{\psi }
\end{equation}%
where $\phi $ and $\mathbf{\psi }=\psi \mathbf{e}_{z}$ are the scalar and
vectorial potentials respectively associated with the longitudinal and
transverse fields. The scalar potentials $\phi $ and $\psi $ satisfy the
Helmholtz equations $\nabla ^{2}\phi +k_{L}^{2}\phi =0$ and $\nabla ^{2}\psi
+k_{T}^{2}\psi =0$. Since the longitudinal and transverse fields in the
elastic elliptical disc have no singularity in the neighbourhood of the
origin, they can be expanded in terms of Bessel functions as 
\begin{equation}
\phi =\sum_{n=-\infty }^{+\infty }A_{n}^{L}J_{n}(k_{L}\rho )e^{in\theta
},\;\;\;\;\;\psi =\sum_{n=-\infty }^{+\infty }A_{n}^{T}J_{n}(k_{T}\rho
)e^{in\theta }
\end{equation}%
where $A_{n}^{j}$ are unknown coefficients with $j\in \{L,T\}$. The physical
quantity to consider in what follows is the elastic displacement expressed
by 
\begin{equation}
\mathbf{u}=u_{\rho }\mathbf{e}_{\rho }+u_{\theta }\mathbf{e}_{\theta
}=\left( -\frac{\partial \phi }{\partial \rho }+\frac{1}{\rho }\frac{%
\partial \psi }{\partial \theta }\right) \mathbf{e}_{\rho }+\left( -\frac{1}{%
\rho }\frac{\partial \phi }{\partial \theta }-\frac{\partial \psi }{\partial
\rho }\right) \mathbf{e}_{\theta }.  \label{displacement}
\end{equation}

\subsection{Symmetry considerations and group theory}

\begin{figure}[!h]
\includegraphics[scale=0.7]{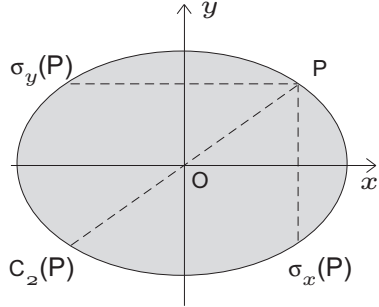}  \centering
\caption{Symmetry transformations.}
\label{fig:transformations}
\end{figure}

We extend to a vectorial formalism a method already used in the context of a
scalar theory \cite{yves, paul, stephane}. The elliptical disc is invariant
under four symmetry transformations (figure \ref{fig:transformations}): (i) $%
E$, the identity transformation $\left( \theta \rightarrow \theta \right) $,
(ii) $C_{2}$, the rotation through $\pi $ about the $Oz$ axis $\left( \theta
\rightarrow \pi +\theta \right) $, (iii) $\sigma _{x}$, the mirror
reflection in the plane $Oxz$ $\left( \theta \rightarrow -\theta \right) $,
(iv) $\sigma _{y}$, the mirror reflection in the plane $Oyz$ $\left( \theta
\rightarrow \pi -\theta \right) $. 
These four transformations form a finite group, called $\mathcal{C}_{2v}$,
which is the symmetry group of the elliptical disc \cite{morton}. The action
of these transformations on the basis vectors $\mathbf{e}_{\rho }$ and $%
\mathbf{e}_{\theta }$ is given by 
\numparts
\begin{eqnarray}
E(\mathbf{e}_{\rho }) &=&\mathbf{e}_{\rho };\quad C_{2}(\mathbf{e}_{\rho })=%
\mathbf{e}_{\rho };\quad\sigma _{x}(\mathbf{e}_{\rho })=\mathbf{e}_{\rho
};\quad\;\;\sigma _{y}(\mathbf{e}_{\rho })=\mathbf{e}_{\rho } \\
E(\mathbf{e}_{\theta }) &=&\mathbf{e}_{\theta };\quad C_{2}(\mathbf{e}%
_{\theta })=\mathbf{e}_{\theta };\quad\sigma _{x}(\mathbf{e}_{\theta })=-%
\mathbf{e}_{\theta };\quad\sigma _{y}(\mathbf{e}_{\theta })=-\mathbf{e}%
_{\theta }
\end{eqnarray}
\endnumparts
Four one-dimensional irreducible representations labelled $A_{1}$, $A_{2}$, $%
B_{1}$, $B_{2}$ are associated with this symmetry group $\mathcal{C}_{2v}$.
In a given representation ($A_{1}$, $A_{2}$, $B_{1}$ or $B_{2}$) the group
elements $E$, $C_{2}$, $\sigma _{x}$ and $\sigma _{y}$ are represented by $%
1\times 1$ matrices given in the corresponding row of the character table
(table \ref{table:character}) 
\Table{\label{table:character}Character table of $\mathcal{C}_{2v}$.}
\br
$\mathcal{C}_{2v}:$ & $E$ & $C_{2}$ & $\sigma _{x}$ & $\sigma _{y}$\\
\mr 
$A_{1}$ & 1 & 1 & 1 & 1 \\
$A_{2}$ & 1 & 1 & \-{1} & \-{1} \\
$B_{1}$ & 1 & \-{1} & 1 & \-{1} \\
$B_{2}$ & 1 & \-{1} & \-{1} & 1 \\
\br
\endTable%
. From the character table \ref{table:character}, we can obviously deduce,
by applying the four symmetry transformations to the basis vectors, that $%
\mathbf{e}_{\rho }$ belongs to the representation $A_{1}$ and $\mathbf{e}%
_{\theta }$ belongs to the representation $A_{2}$.

Let us consider a vectorial function $\mathbf{f}(\rho ,\theta )$ expressed
in the polar coordinates system as 
\begin{equation}
\mathbf{f}(\rho ,\theta )=f_{\rho }\mathbf{e}_{\rho }+f_{\theta }\mathbf{e}%
_{\theta }.
\end{equation}%
The character table permits one to split the scalar components $f_{\rho
},f_{\theta }$ as a sum of functions belonging to the four irreducible
representations of $\mathcal{C}_{2v}$. Indeed, one can write 
\begin{equation}
f_{i}(\rho ,\theta )=f_{i}^{A_{1}}(\rho ,\theta )+f_{i}^{A_{2}}(\rho ,\theta
)+f_{i}^{B_{1}}(\rho ,\theta )+f_{i}^{B_{2}}(\rho ,\theta )\;_{;}\;i\in
\{\rho ,\theta \}
\end{equation}%
with $f_{i}^{A_{1}}(\rho ,\theta )$, $f_{i}^{A_{2}}(\rho ,\theta )$, $%
f_{i}^{B_{1}}(\rho ,\theta )$, $f_{i}^{B_{2}}(\rho ,\theta )$ satisfying 
\numparts
\begin{eqnarray}
Ef_{i}^{A_{1}} &=&f_{i}^{A_{1}};\quad C_{2}f_{i}^{A_{1}}=f_{i}^{A_{1}};\quad%
\;\;\sigma _{x}f_{i}^{A_{1}}=f_{i}^{A_{1}};\quad\;\;\sigma
_{y}f_{i}^{A_{1}}=f_{i}^{A_{1}}  \label{actionECsigmaA1} \\
Ef_{i}^{A_{2}} &=&f_{i}^{A_{2}};\quad C_{2}f_{i}^{A_{2}}=f_{i}^{A_{2}};\quad%
\;\;\sigma _{x}f_{i}^{A_{2}}=-f_{i}^{A_{2}};\quad\sigma
_{y}f_{i}^{A_{2}}=-f_{i}^{A_{2}}  \label{actionECsigmaA2} \\
Ef_{i}^{B_{1}} &=&f_{i}^{B_{1}};\quad C_{2}f_{i}^{B_{1}}=-f_{i}^{B_{1}};\quad%
\sigma _{x}f_{i}^{B_{1}}=f_{i}^{B_{1}};\quad\;\;\sigma
_{y}f_{i}^{B_{1}}=-f_{i}^{B_{1}}  \label{actionECsigmaB1} \\
Ef_{i}^{B_{2}} &=&f_{i}^{B_{2}};\quad C_{2}f_{i}^{B_{2}}=-f_{i}^{B_{2}};\quad%
\sigma _{x}f_{i}^{B_{2}}=-f_{i}^{B_{2}};\quad\sigma
_{y}f_{i}^{B_{2}}=f_{i}^{B_{2}}.  \label{actionECsigmaB2}
\end{eqnarray}
\endnumparts
Hence, the vectorial function can be expressed as 
\begin{equation}
\mathbf{f}(\rho ,\theta )=\left( f_{\rho }^{A_{1}}+f_{\rho }^{A_{2}}+f_{\rho
}^{B_{1}}+f_{\rho }^{B_{2}}\right) \mathbf{e}_{\rho }+\left( f_{\theta
}^{A_{1}}+f_{\theta }^{A_{2}}+f_{\theta }^{B_{1}}+f_{\theta }^{B_{2}}\right) 
\mathbf{e}_{\theta },  \label{vector decomp}
\end{equation}%
where the symmetry decomposition over $A_{1},$ $A_{2},$ $B_{1},$ $B_{2}$ is
only carried out for the scalar components.

Since group theory will be applied to the displacement vector $\mathbf{u,}$
we need to generalize the decomposition over the four irreducible
representations in a purely vector context. Expanding equation (\ref{vector
decomp}), by taking into account that $\mathbf{e}_{\rho }\in A_{1}$ and $%
\mathbf{e}_{\theta }\in A_{2},$ and according to the multiplication table %
\ref{table:multiplication}, one can write $\mathbf{f}(\rho ,\theta )$ as a
sum of vector functions, each belonging to a given irreducible
representation of $\mathcal{C}_{2v},$ in the form 
\begin{equation}
\mathbf{f}(\rho ,\theta )=\mathbf{f}^{A_{1}}(\rho ,\theta )+\mathbf{f}%
^{A_{2}}(\rho ,\theta )+\mathbf{f}^{B_{1}}(\rho ,\theta )+\mathbf{f}%
^{B_{2}}(\rho ,\theta ),
\end{equation}%
with 
\numparts
\begin{eqnarray}
\mathbf{f}^{A_{1}}& =&f_{\rho }^{A_{1}}\mathbf{e}_{\rho }+f_{\theta }^{A_{2}}%
\mathbf{e}_{\theta }  \label{vectorielA1} \\
\mathbf{f}^{A_{2}}& =&f_{\rho }^{A_{2}}\mathbf{e}_{\rho }+f_{\theta }^{A_{1}}%
\mathbf{e}_{\theta }  \label{vectorielA2} \\
\mathbf{f}^{B_{1}}& =&f_{\rho }^{B_{1}}\mathbf{e}_{\rho }+f_{\theta }^{B_{2}}%
\mathbf{e}_{\theta }  \label{vectorielB1} \\
\mathbf{f}^{B_{2}}& =&f_{\rho }^{B_{2}}\mathbf{e}_{\rho }+f_{\theta }^{B_{1}}%
\mathbf{e}_{\theta }.  \label{vectorielB2}
\end{eqnarray}%
\endnumparts
It should be noted that the scalar components of a vectorial function in a
given irreducible representation do not necessarily belong to the same
representation (see (\ref{vectorielA1}-\ref{vectorielB2})).%
\Table{\label{table:multiplication}Multiplication table for a function belonging to a given irreducible representation.}
\br
$\times$ & $A_{1}$ & $A_{2}$ & $B_{1}$ & $B_{2}$\\
\mr 
$A_{1}$ & $A_{1}$ & $A_{2}$ & $B_{1}$ & $B_{2}$ \\
$A_{2}$ & $A_{2}$ & $A_{1}$ & $B_{2}$ & $B_{1}$ \\
$B_{1}$ & $B_{1}$ & $B_{2}$ & $A_{1}$ & $A_{2}$ \\
$B_{2}$ & $B_{2}$ & $B_{1}$ & $A_{2}$ & $A_{1}$ \\
\br
\endTable%

\begin{figure}[h]
\includegraphics[scale=0.8]{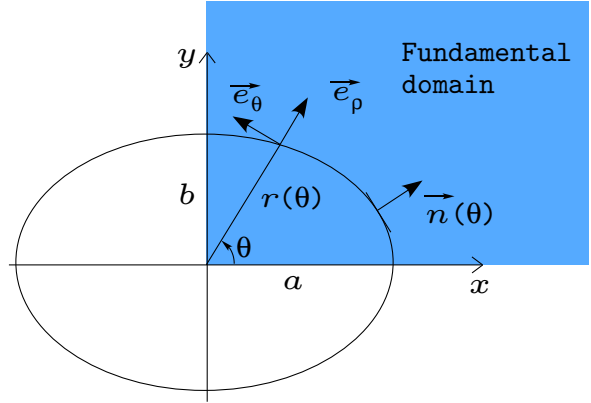}  \centering
\caption{Fundamental domain.}
\label{fig:fondamental}
\end{figure}

The use of group theory allows us to restrict the study to the so-called
fundamental domain which reduces to a quarter of the elliptical disc $\theta
\in \left[ 0,\pi /2\right] $ (see figure \ref{fig:fondamental}). This
constitutes a great improvement from both theoretical and numerical point of
view. Of course, the physical quantities of interest can be determined for
the full domain from simple symmetry considerations (table \ref%
{table:character}). 

\subsection{Solving the eigenvalue problem}

The eigenfrequencies are calculated by applying the traction-free boundary
condition $\mathbf{t}=\overline{\overline{\mathbf{\sigma }}}\cdot \mathbf{n}%
(\theta )=\mathbf{0}$ on the ellipse $\rho =r(\theta )$, i.e. 
\begin{equation}
t_{\rho }\mathbf{e}_{\rho }+t_{\theta }\mathbf{e}_{\theta }=\mathbf{0}
\label{boundary cond}
\end{equation}%
with, according to equation (\ref{normale}), 
\numparts
\begin{eqnarray}
t_{\rho } &=&\sigma _{\rho \rho }n_{\rho }+\sigma _{\rho \theta }n_{\theta },
\label{trho} \\
t_{\theta } &=&\sigma _{\rho \theta }n_{\rho }+\sigma _{\theta \theta
}n_{\theta }.  \label{ttheta}
\end{eqnarray}%
\endnumparts
The components of $\overline{\overline{\mathbf{\sigma }}}$ in circular
cylindrical coordinates are obtained from the strain tensor \cite{landau},
Hooke's law, and the Helmholtz equation for the scalar potential $\phi $ 
\numparts
\begin{eqnarray}
\sigma _{\rho \rho }& =&\lambda k_{L}^{2}\phi -2\mu \biggl(\frac{\partial
^{2}\phi }{\partial \rho ^{2}}+\frac{1}{\rho ^{2}}\frac{\partial \psi }{%
\partial \theta }-\frac{1}{\rho }\frac{\partial ^{2}\psi }{\partial \rho
\partial \theta }\biggr), \\
\sigma _{\rho \theta }& =&\sigma _{\theta \rho }=\mu \biggl(-\frac{2}{\rho }%
\frac{\partial ^{2}\phi }{\partial \theta \partial \rho }+\frac{1}{\rho ^{2}}%
\frac{\partial ^{2}\psi }{\partial \theta ^{2}}+\frac{1}{\rho }\frac{%
\partial \psi }{\partial \rho }-\frac{\partial ^{2}\psi }{\partial \rho ^{2}}%
+\frac{2}{\rho ^{2}}\frac{\partial \phi }{\partial \theta }\biggr), \\
\sigma _{\theta \theta }& =&\lambda k_{L}^{2}\phi +2\mu \biggl(-\frac{1}{%
\rho ^{2}}\frac{\partial ^{2}\phi }{\partial \theta ^{2}}-\frac{1}{\rho }%
\frac{\partial ^{2}\psi }{\partial \theta \partial \rho }-\frac{1}{\rho }%
\frac{\partial \phi }{\partial \rho }+\frac{1}{\rho ^{2}}\frac{\partial \psi 
}{\partial \theta }\biggr).
\end{eqnarray}
\endnumparts
The boundary condition (\ref{boundary cond}) can now be expressed separately
in each irreducible representation. We note that $\mathbf{n}(\theta )$
defined by (\ref{normale}) belongs to $A_{1}$ for obvious symmetry
considerations, thus its components $n_{\rho }$ and $n_{\theta }$
respectively belong to $A_{1}$ and $A_{2}$ under the property (\ref%
{vectorielA1}). In what follows, we use the latter property, the
multiplication table (table \ref{table:multiplication}) and we note that the
derivative by respect to $\theta $ acts as a multiplication by some function
belonging to the representation $A_{2}$.

For clarity, we also define 
\begin{equation}
\sum_{n\;\mathit{even}}f_{n}:=\sum_{r=0}^{+\infty }f_{2r},\;\;\;\mathit{and}%
\;\sum_{n\;\mathit{odd}}f_{n}:=\sum_{r=0}^{+\infty }f_{2r+1}.
\end{equation}

\subsubsection{Representation $A_{1}$}

We express the boundary condition $\mathbf{t}=\mathbf{0}$ in the irreducible
representation $A_{1}$, using (\ref{vectorielA1}) with $\phi ^{A_{1}}$ and $%
\psi ^{A_{2}}$ satisfying (\ref{actionECsigmaA1}) and (\ref{actionECsigmaA2}%
) given by 
\begin{equation}
\fl\phi ^{A_{1}}(\rho ,\theta )=\sum_{n\;\mathit{even}}\gamma
_{n}A_{n}^{L}J_{n}(k_{L}\rho )\cos n\theta ,\;\;\;\;\;\psi ^{A_{2}}(\rho
,\theta )=\sum_{n\;\mathit{even}}\gamma _{n}A_{n}^{T}J_{n}(k_{T}\rho )\sin
n\theta ,  \label{phiA1_psiA2}
\end{equation}%
where $\gamma _{n}$ is the Neumann factor defined by $\gamma _{0}=1$ and $%
\gamma _{n}=2$ for $n>0.$ We obtain 
\numparts%
\begin{eqnarray}
\fl%
\sum_{n\;\mathit{even}}\gamma _{n}\left[ A_{n}^{L}\left( c_{n}^{L}n_{\rho
}\cos n\theta +d_{n}^{L}n_{\theta }\sin n\theta \right) +A_{n}^{T}\left(
c_{n}^{T}n_{\rho }\cos n\theta +d_{n}^{T}n_{\theta }\sin n\theta \right) %
\right]  &=&0  \label{BCA1step1a} \\
\fl%
\sum_{n\;\mathit{even}}\gamma _{n}\left[ A_{n}^{L}\left( d_{n}^{L}n_{\rho
}\sin n\theta -f_{n}^{L}n_{\theta }\cos n\theta \right) +A_{n}^{T}\left(
d_{n}^{T}n_{\rho }\sin n\theta -f_{n}^{T}n_{\theta }\cos n\theta \right) %
\right]  &=&0  \label{BCA1step1b}
\end{eqnarray}%
\endnumparts where the structural functions $c_{n}^{j},$ $d_{n}^{j},$ $%
f_{n}^{j},$ $j\in \{L,T\}$ are given in \ref{coefficients}. To overcome the
angular dependence, the functions $F_{n}(\theta )$ in parenthesis appearing
in (\ref{BCA1step1a}) and (\ref{BCA1step1b}) are expanded in Fourier series
by setting 
\begin{equation}
F_{n}(\theta )=\sum_{p=-\infty }^{+\infty }f_{n,p}e^{ip\theta },\;\;\;\mathit{
with}\;f_{n,p}=\frac{1}{2\pi }\int_{-\pi }^{\pi }F_{n}(\theta )e^{-ip\theta
}d\theta .  \label{fourierseries}
\end{equation}%
In (\ref{fourierseries}), the restriction to the fundamental domain and the
parity of $F_{n}(\theta )$ permits one to reduce the domain of integration
from $\left[ -\pi ,\pi \right] $ to $\left[ 0,\pi /2\right] $ and the sum
over $p$ from $\left[ -\infty ,+\infty \right] $ to $\left[ 0,+\infty \right]
$ with $p$ even. Then (\ref{BCA1step1a}) and (\ref{BCA1step1b}) lead to the
final equations 
\numparts%
\begin{eqnarray}
&&\sum_{n\;\mathit{even}}\sum_{p\;\mathit{even}}\gamma _{n}\gamma _{p}\left(
A_{n}^{L}\,\beta _{n,p}^{L}+A_{n}^{T}\,\beta _{n,p}^{T}\right) \cos p\theta
=0  \label{resulteqnA1a} \\
&&\sum_{n\;\mathit{even}}\sum_{p\;\mathit{even}}\gamma _{n}\left(
A_{n}^{L}\,\xi _{n,p}^{L}+A_{n}^{T}\,\xi _{n,p}^{T}\right) \sin p\theta =0
\label{resulteqnA1b}
\end{eqnarray}%
\endnumparts where the Fourier coefficients $\beta _{n,p}^{j},\xi _{n,p}^{j}$%
, $j\in \{L,T\}$ are given in \ref{fouriercoeff}.

Proceeding by a similar way, we obtain \textit{mutatis mutandis} the final
equations for the representations $A_{2},$ $B_{1}$ and $B_{2}.$

\subsubsection{Representation $A_{2}$}

Using (\ref{vectorielA2}), $\phi ^{A_{2}}$ and $\psi ^{A_{1}}$ satisfying (%
\ref{actionECsigmaA1}) and (\ref{actionECsigmaA2}) given by 
\begin{equation}
\fl\phi ^{A_{2}}(\rho ,\theta )=\sum_{n\;\mathit{even}}\gamma
_{n}A_{n}^{L}J_{n}(k_{L}\rho )\sin n\theta ,\;\;\;\;\;\psi ^{A_{1}}(\rho
,\theta )=\sum_{n\;\mathit{even}}\gamma _{n}A_{n}^{T}J_{n}(k_{T}\rho )\cos
n\theta ,  \label{phiA2_psiA1}
\end{equation}%
the boundary condition reads 
\numparts%
\begin{eqnarray}
\fl%
\sum_{n\;\mathit{even}}\gamma _{n}\left[ A_{n}^{L}\left( c_{n}^{L}n_{\rho
}\sin n\theta -d_{n}^{L}n_{\theta }\cos n\theta \right) +A_{n}^{T}\left(
-c_{n}^{T}n_{\rho }\sin n\theta +d_{n}^{T}n_{\theta }\cos n\theta \right) %
\right]  &=&0  \label{BCA2step1a} \\
\fl%
\sum_{n\;\mathit{even}}\gamma _{n}\left[ A_{n}^{L}\left( -d_{n}^{L}n_{\rho
}\cos n\theta -f_{n}^{L}n_{\theta }\sin n\theta \right) +A_{n}^{T}\left(
d_{n}^{T}n_{\rho }\cos n\theta +f_{n}^{T}n_{\theta }\sin n\theta \right) %
\right]  &=&0  \label{BCA2step1b}
\end{eqnarray}%
\endnumparts where the structural functions $c_{n}^{j},$ $d_{n}^{j},$ $%
f_{n}^{j},$ $j\in \{L,T\}$ are given in \ref{coefficients}. Then (\ref%
{BCA2step1a}) and (\ref{BCA2step1b}) lead to the final equations 
\numparts%
\begin{eqnarray}
&&\sum_{n\;\mathit{even}}\sum_{p\;\mathit{even}}\gamma _{n}\left(
A_{n}^{L}\,\alpha _{n,p}^{L}+A_{n}^{T}\,\alpha _{n,p}^{T}\right) \sin
p\theta =0  \label{resulteqnA2a} \\
&&\sum_{n\;\mathit{even}}\sum_{p\;\mathit{even}}\gamma _{n}\gamma _{p}\left(
A_{n}^{L}\,\eta _{n,p}^{L}+A_{n}^{T}\,\eta _{n,p}^{T}\right) \cos p\theta =0
\label{resulteqnA2b}
\end{eqnarray}%
\endnumparts where the Fourier coefficients $\alpha _{n,p}^{j},\eta
_{n,p}^{j}$, $j\in \{L,T\}$ are given in \ref{fouriercoeff}.

\subsubsection{Representation $B_{1}$}

Using (\ref{vectorielB1}), $\phi ^{B_{1}}$ and $\psi ^{B_{2}}$ satisfying (%
\ref{actionECsigmaB1}) and (\ref{actionECsigmaB2}) given by 
\begin{equation}
\fl\phi ^{B_{1}}(\rho ,\theta )=\sum_{n\;\mathit{odd}}A_{n}^{L}J_{n}(k_{L}%
\rho )\cos n\theta ,\;\;\;\;\;\psi ^{B_{2}}(\rho ,\theta )=\sum_{n\;\mathit{%
odd}}A_{n}^{T}J_{n}(k_{T}\rho )\sin n\theta ,  \label{phiB1_psiB2}
\end{equation}%
the boundary condition reads 
\numparts%
\begin{eqnarray}
\fl%
\sum_{n\;\mathit{odd}}\left[ A_{n}^{L}\left( c_{n}^{L}n_{\rho }\cos n\theta
+d_{n}^{L}n_{\theta }\sin n\theta \right) +A_{n}^{T}\left( c_{n}^{T}n_{\rho
}\cos n\theta +d_{n}^{T}n_{\theta }\sin n\theta \right) \right]  &=&0
\label{BCB1step1a} \\
\fl%
\sum_{n\;\mathit{odd}}\left[ A_{n}^{L}\left( d_{n}^{L}n_{\rho }\sin n\theta
-f_{n}^{L}n_{\theta }\cos n\theta \right) +A_{n}^{T}\left( d_{n}^{T}n_{\rho
}\sin n\theta -f_{n}^{T}n_{\theta }\cos n\theta \right) \right]  &=&0
\label{BCB1step1b}
\end{eqnarray}%
\endnumparts where the structural functions $c_{n}^{j},$ $d_{n}^{j},$ $%
f_{n}^{j},$ $j\in \{L,T\}$ are given in \ref{coefficients}. Then (\ref%
{BCB1step1a}) and (\ref{BCB1step1b}) lead to the final equations 
\numparts%
\begin{eqnarray}
&&\sum_{n\;\mathit{odd}}\sum_{p\;\mathit{odd}}\left( A_{n}^{L}\,\beta
_{n,p}^{L}+A_{n}^{T}\,\beta _{n,p}^{T}\right) \cos p\theta =0
\label{resulteqnB1a} \\
&&\sum_{n\;\mathit{odd}}\sum_{p\;\mathit{odd}}\left( A_{n}^{L}\,\xi
_{n,p}^{L}+A_{n}^{T}\,\xi _{n,p}^{T}\right) \sin p\theta =0
\label{resulteqnB1b}
\end{eqnarray}%
\endnumparts where the Fourier coefficients $\beta _{n,p}^{j},\xi _{n,p}^{j}$%
, $j\in \{L,T\}$ are given in \ref{fouriercoeff}.

\subsubsection{Representation $B_{2}$}

Using (\ref{vectorielB2}), $\phi ^{B_{2}}$ and $\psi ^{B_{1}}$ satisfying (%
\ref{actionECsigmaB1}) and (\ref{actionECsigmaB2}) given by 
\begin{equation}
\fl\phi ^{B_{2}}(\rho ,\theta )=\sum_{n\;\mathit{odd}}A_{n}^{L}J_{n}(k_{L}%
\rho )\sin n\theta ,\;\;\;\;\;\psi ^{B_{1}}(\rho ,\theta )=\sum_{n\;\mathit{%
odd}}A_{n}^{T}J_{n}(k_{T}\rho )\cos n\theta ,  \label{phiB2_psiB1}
\end{equation}%
the boundary condition reads 
\numparts%
\begin{eqnarray}
\fl%
\sum_{n\;\mathit{odd}}\left[ A_{n}^{L}\left( c_{n}^{L}n_{\rho }\sin n\theta
-d_{n}^{L}n_{\theta }\cos n\theta \right) +A_{n}^{T}\left( -c_{n}^{T}n_{\rho
}\sin n\theta +d_{n}^{T}n_{\theta }\cos n\theta \right) \right]  &=&0
\label{BCB2step1a} \\
\fl%
\sum_{n\;\mathit{odd}}\left[ A_{n}^{L}\left( -d_{n}^{L}n_{\rho }\cos n\theta
-f_{n}^{L}n_{\theta }\sin n\theta \right) +A_{n}^{T}\left( d_{n}^{T}n_{\rho
}\cos n\theta +f_{n}^{T}n_{\theta }\sin n\theta \right) \right]  &=&0
\label{BCB2step1b}
\end{eqnarray}%
\endnumparts where the structural functions $c_{n}^{j},$ $d_{n}^{j},$ $%
f_{n}^{j},$ $j\in \{L,T\}$ are given in \ref{coefficients}. Then (\ref%
{BCB2step1a}) and (\ref{BCB2step1b}) lead to the final equations 
\numparts%
\begin{eqnarray}
&&\sum_{n\;\mathit{odd}}\sum_{p\;\mathit{odd}}\left( A_{n}^{L}\,\alpha
_{n,p}^{L}+A_{n}^{T}\,\alpha _{n,p}^{T}\right) \sin p\theta =0
\label{resulteqnB2a} \\
&&\sum_{n\;\mathit{odd}}\sum_{p\;\mathit{odd}}\left( A_{n}^{L}\,\eta
_{n,p}^{L}+A_{n}^{T}\,\eta _{n,p}^{T}\right) \cos p\theta =0
\label{resulteqnB2b}
\end{eqnarray}%
\endnumparts where the Fourier coefficients $\alpha _{n,p}^{j},\eta
_{n,p}^{j}$, $j\in \{L,T\}$ are given in \ref{fouriercoeff}.

\subsubsection{Matrix form}

\label{matrixform}

The above systems of equations (\ref{resulteqnA1a}-\ref{resulteqnA1b}), (\ref%
{resulteqnA2a}-\ref{resulteqnA2b}),  (\ref{resulteqnB1a}-\ref{resulteqnB1b})
and (\ref{resulteqnB2a}-\ref{resulteqnB2b}) can be expressed in matrix form,
for each representation $R_{i}$, by defining a block matrix 
\begin{equation}
M^{R_{i}}=\left( 
\begin{array}{cc}
P^{R_{i}L} & P^{R_{i}T} \\ 
&  \\ 
Q^{R_{i}L} & Q^{R_{i}T}%
\end{array}%
\right) ,  \label{matrix}
\end{equation}%
where $(P,Q)\in \{(\beta _{q,r}^{j},\xi _{q,r}^{j}),(\alpha _{q,r}^{j},\eta
_{q,r}^{j})\}$, the Fourier coefficients given in \ref{fouriercoeff}. Then,
the resonances are determined by solving the characteristic equation 
\begin{equation}
\det \;M^{R_{i}}\;=0.  \label{eq_caract}
\end{equation}

\subsection{Elastic displacement}

The elastic displacement can be expressed in each irreducible representation 
$R_{i}$ using (\ref{displacement}) and (\ref{vectorielA1}-\ref{vectorielB2})
and the potentials (\ref{phiA1_psiA2}), (\ref{phiA2_psiA1}), (\ref%
{phiB1_psiB2}) and (\ref{phiB2_psiB1}). Then we get 
\numparts
\label{displacement_in_rep} 
\begin{eqnarray}
\mathbf{u}^{A_{1}}(\rho ,\theta )=\sum_{n\;\mathit{even}}\gamma _{n} &&%
\Biggl[\left( -k_{L}A_{n}^{L}J_{n}^{\prime }(k_{L}\rho )+\frac{n}{\rho }%
A_{n}^{T}J_{n}(k_{T}\rho )\right) \cos n\theta \;\mathbf{e_{\rho }} 
\nonumber \\
&&+\left( -k_{T}A_{n}^{T}J_{n}^{\prime }(k_{T}\rho )+\frac{n}{\rho }%
A_{n}^{L}J_{n}(k_{L}\rho )\right) \sin n\theta \;\mathbf{e_{\theta }}\Biggr]
\\
\mathbf{u}^{A_{2}}(\rho ,\theta )=\sum_{n\;\mathit{even}}\gamma _{n} &&%
\Biggl[\left( -k_{L}A_{n}^{L}J_{n}^{\prime }(k_{L}\rho )-\frac{n}{\rho }%
A_{n}^{T}J_{n}(k_{T}\rho )\right) \sin n\theta \;\mathbf{e_{\rho }} 
\nonumber \\
&&+\left( -k_{T}A_{n}^{T}J_{n}^{\prime }(k_{T}\rho )-\frac{n}{\rho }%
A_{n}^{L}J_{n}(k_{L}\rho )\right) \cos n\theta \;\mathbf{e_{\theta }}\Biggr]
\\
\mathbf{u}^{B_{1}}(\rho ,\theta )=\sum_{n\;\mathit{odd}} &&\Biggl[\left(
-k_{L}A_{n}^{L}J_{n}^{\prime }(k_{L}\rho )+\frac{n}{\rho }%
A_{n}^{T}J_{n}(k_{T}\rho )\right) \cos n\theta \;\mathbf{e_{\rho }} 
\nonumber \\
&&+\left( -k_{T}A_{n}^{T}J_{n}^{\prime }(k_{T}\rho )+\frac{n}{\rho }%
A_{n}^{L}J_{n}(k_{L}\rho )\right) \sin n\theta \;\mathbf{e_{\theta }}\Biggr]
\\
\mathbf{u}^{B_{2}}(\rho ,\theta )=\sum_{n\;\mathit{odd}} &&\Biggl[\left(
-k_{L}A_{n}^{L}J_{n}^{\prime }(k_{L}\rho )-\frac{n}{\rho }%
A_{n}^{T}J_{n}(k_{T}\rho )\right) \sin n\theta \;\mathbf{e_{\rho }} 
\nonumber \\
&&+\left( -k_{T}A_{n}^{T}J_{n}^{\prime }(k_{T}\rho )-\frac{n}{\rho }%
A_{n}^{L}J_{n}(k_{L}\rho )\right) \cos n\theta \;\mathbf{e_{\theta }}\Biggr].
\end{eqnarray}%
\endnumparts
The vibrational normal modes displacement can then be obtained by computing $%
|\mathbf{u}^{R_{i}}|$.

\section{Numerical Results}

\subsection{Splitting up of resonances}

The main advantage of using group theory is to obtain uncoupled equations
that can be solved separately for each irreducible representation. As
explained in section \ref{matrixform}, the resonances are determined by
solving the characteristic equation $\det \;M^{R_{i}}\;=0$ where $M^{R_{i}}$
is defined by (\ref{matrix}). In what follows, the $k$-resonances are
normalized by defining the refractive indices $c_{w}/c_{L}$ and $c_{w}/c_{T}$%
, with $c_{w}=1480$ m.s$^{-1}$ the sound velocity in water. The same
normalization will be used in the further study of acoustic scattering by
infinite cylinders of elliptical cross-section (work in progress).

Since these matrices are of infinite dimensionality, some truncation must be
performed in order to generate a numerical solution. The chosen truncation
order $N$ depends on the dimensionless reduced wave number $ka$ ($a$ is the
semi-major axis of the elliptical disc) and it has been numerically
investigated. This method improves the matrix conditioning and the speed of
computations. Indeed, the matrices involved in our problem are of dimension $%
\left( N\times N\right) $ instead of $\left( 2N\times 2N\right) $ for the
coupled problem (without group theory).

\begin{figure}[]
\includegraphics[scale=0.5]{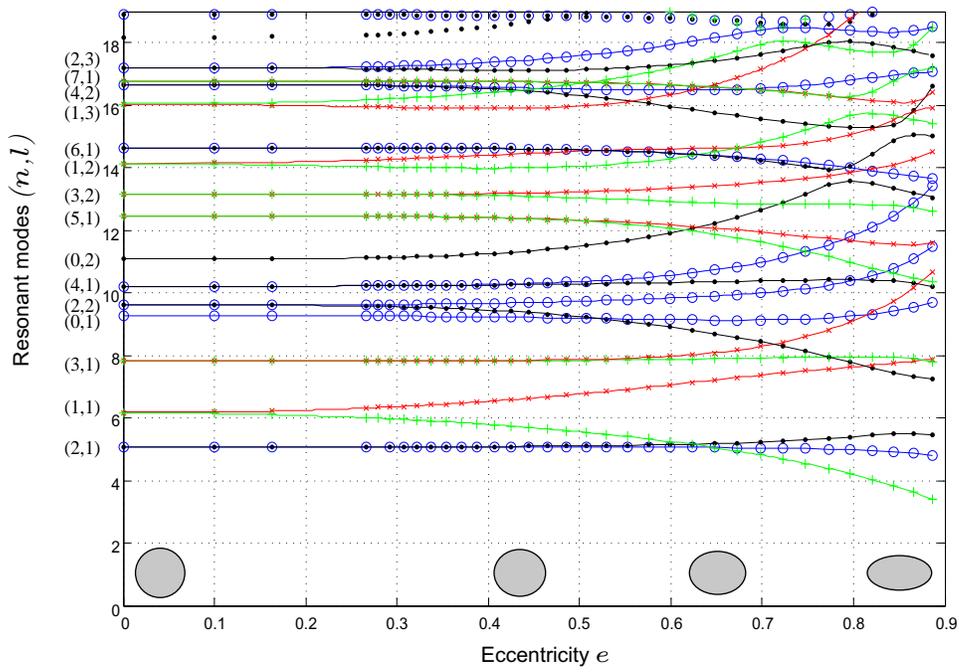}  \centering
\caption{Tracking of the $k$-resonances versus the eccentricity $e$: ($\circ
,A_{1}$) ; ($\cdot ,A_{2}$) ; $(\times ,B_{1})$ ; $(+,B_{2})$.}
\label{fig:splitting}
\end{figure}

The problem is uncoupled over the four irreducible representations of $%
\mathcal{C}_{2v}$ and this provides a full classification of the resonances.
In figure \ref{fig:splitting}, the $k$-resonances associated with the first
fifteen resonant modes labelled by $\left( n,\ell \right) $ in the usual
modal formalism are plotted as the circular disc is deformed to the
elliptical one, by keeping the mean radius constant. The resonances are
tagged and tracked as the eccentricity increases and two important effects
are observed: the splitting up of resonances and the crossings between
resonant modes. The splitting up can be interpreted in terms of symmetry
breaking $\mathcal{O}(2)\rightarrow \mathcal{C}_{2v}$. Moreover,

\begin{itemize}
\item[-] for any even angular index $n$, the resonances are splitted in the $%
A_{1}$ and $A_{2}$ irreducible representations (figure \ref%
{fig:splittingeven}),

\item[-] for any odd angular index $n$, the resonances are splitted in the $%
B_{1} $ and $B_{2}$ irreducible representations (figure \ref%
{fig:splittingodd}).
\end{itemize}

\begin{figure}[]
\includegraphics[scale=0.5]{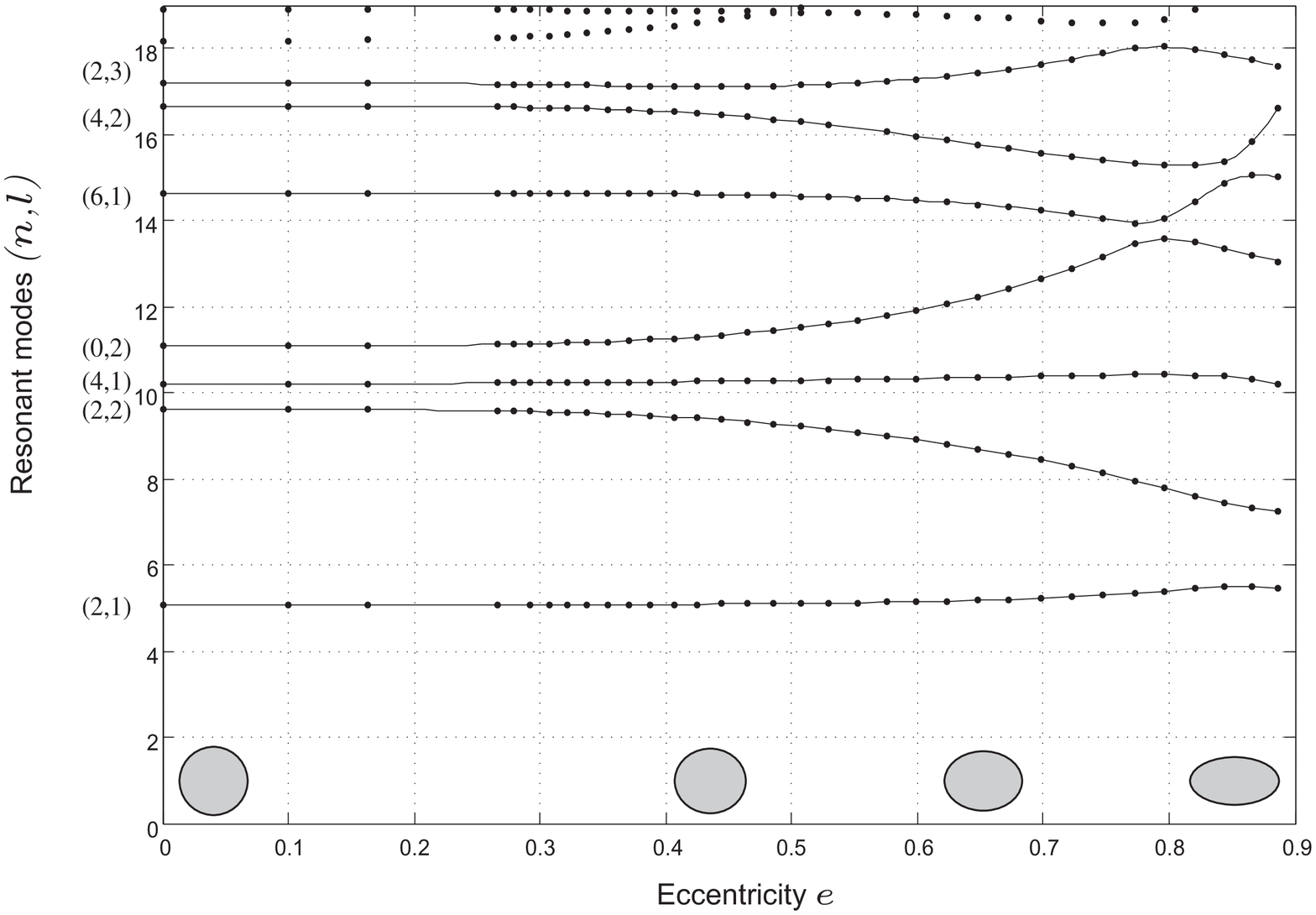}  \centering
\caption{Mode interaction.}
\label{fig:interaction}
\end{figure}

There is no interaction between the resonant modes belonging to different
irreducible representations. Nevertheless, we observe that the resonant
modes belonging to the same representation can interact (repelling modes) as
displayed in figure \ref{fig:interaction} for the modes $\left( 0,2\right) $
and $\left( 6,1\right) $. 

\begin{figure}[]
\includegraphics[scale=0.5]{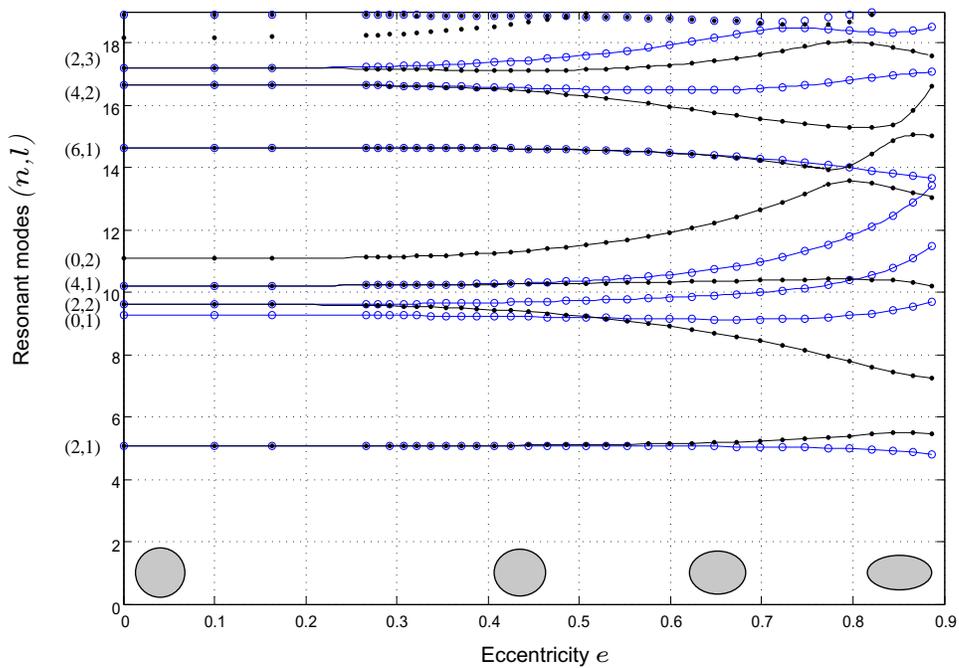}  \centering
\caption{Splitting $(\times ,A_{1})$ ; $(+,A_{2})$, even angular index $n.$}
\label{fig:splittingeven}
\end{figure}

\begin{figure}[]
\includegraphics[scale=0.5]{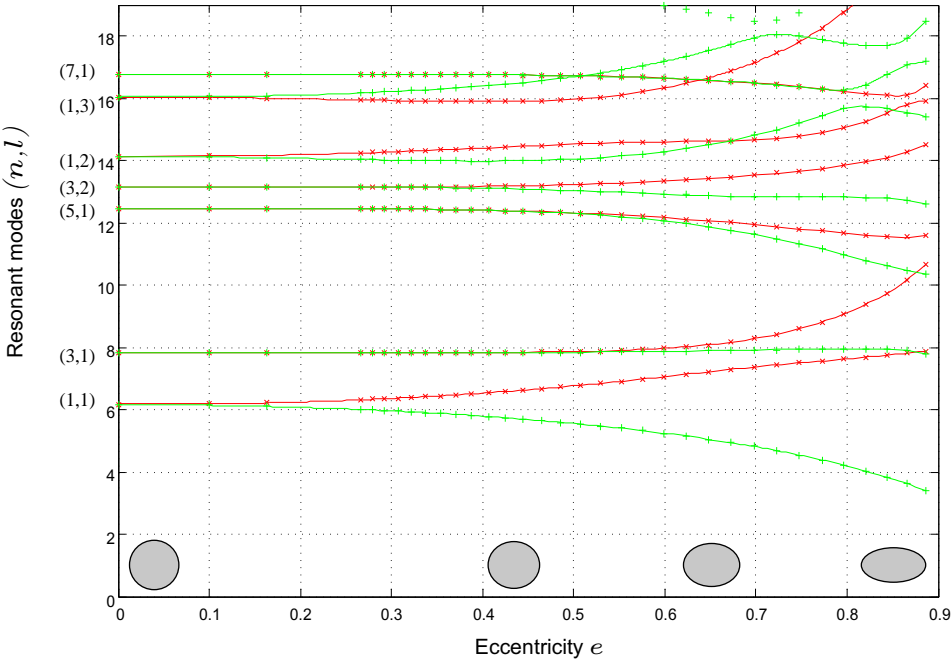}  \centering
\caption{Splitting $(\times ,B_{1})$ ; $(+,B_{2})$, odd angular index $n.$}
\label{fig:splittingodd}
\end{figure}

\subsection{Vibrational normal modes displacement}

\begin{figure}[h]
\includegraphics[scale=0.8]{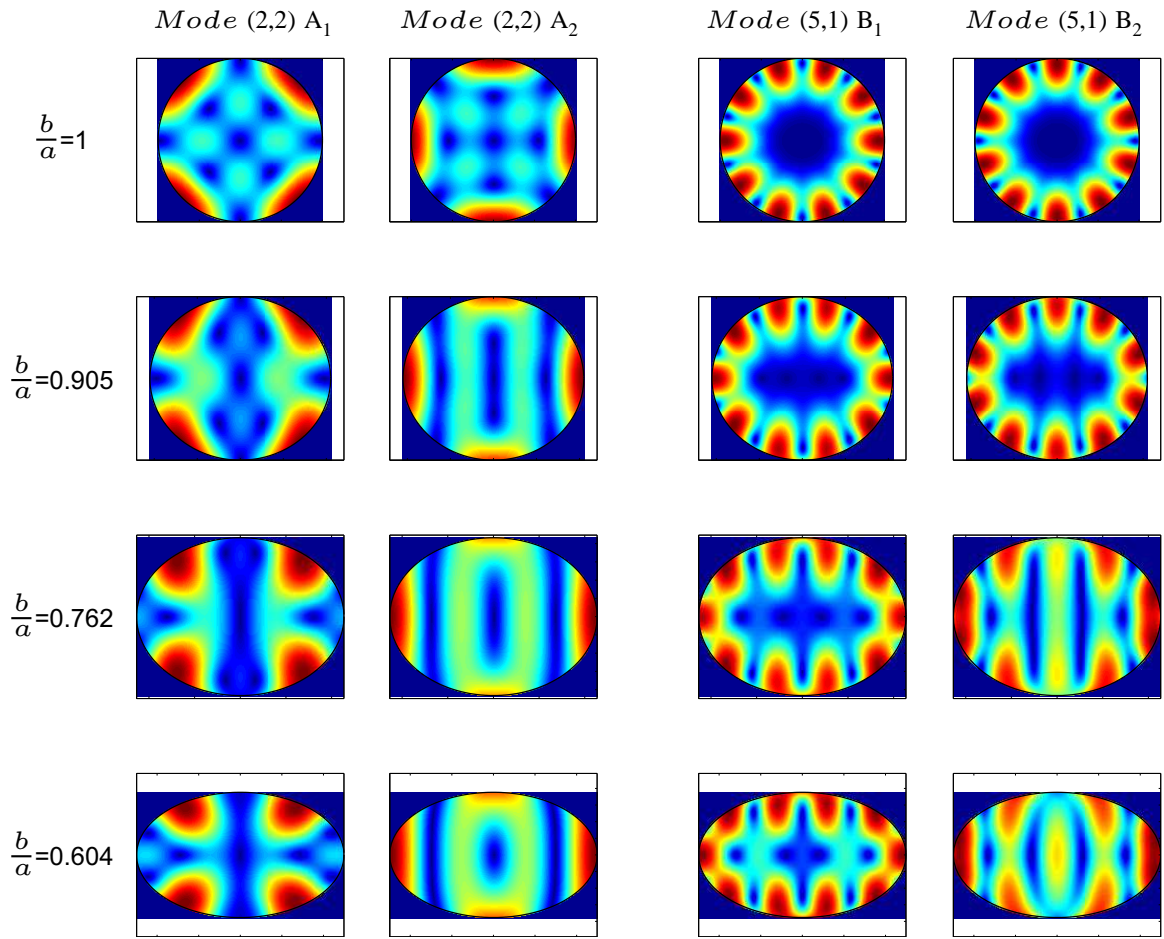}  \centering
\caption{Vibrational normal modes displacement.}
\label{fig:splitting_modes}
\end{figure}

The elastic displacement associated with each irreducible representation $%
R_{i}$ permits to compute the vibrational normal modes displacement. We plot 
$\left\vert \mathbf{u}^{R_{i}}\right\vert $ for various eccentricities. This
computation requires the evaluation of the unknown coefficients to determine
the fields $\phi $ and $\psi $ involved in (\ref{displacement_in_rep}).
These coefficients are calculated by solving the systems of equations (\ref%
{resulteqnA1a}-\ref{resulteqnA1b}), (\ref{resulteqnA2a}-\ref{resulteqnA2b}),
(\ref{resulteqnB1a}-\ref{resulteqnB1b}) and (\ref{resulteqnB2a}-\ref%
{resulteqnB2b}).  In the case of circular disc, for a given degenerated
mode, the transition from a representation to an other can be done by a
simple rotation of an angle $\pi /2n$ (see figure \ref{fig:splitting_modes}
for $b/a=1$). Once the circular disc is deformed into the elliptical one,
the symmetry is broken and patterns become quite different. Mode splittings
are observed as the eccentricity increases (figure \ref{fig:splitting_modes}%
). Only group theory permits one to highlight the splitting up and to link
the splitted modes with the initial degenerated mode of the circular disc.

\FloatBarrier

\subsection{Periodic orbits}

In order to give a physical interpretation in terms of trajectories, we use
periodic orbits theory \cite{berrytabor, tabor}. We numerically compute the
spectral counting function 
\begin{equation}
\mathcal{N}(k)=\sum_{n}\Theta (k-k_{n})  \label{countingfunction}
\end{equation}%
giving the number of the $k$-resonances. The equation (\ref{countingfunction}%
) can be quite generally written in terms of a smooth part and an
oscillating part, the latter containing the periodic orbits contribution,
that is 
\begin{equation}
\mathcal{N}(k)=\mathcal{N}_{\mathit{smooth}}(k)+\mathcal{N}_{\mathit{oscill}}(k).
\label{countingfunctiondecompo}
\end{equation}%
General results for the smooth part of the counting function of isotropic
elastic media with free boundary conditions can be found in \cite%
{dupuismazo, vasilev}. The decomposition (\ref{countingfunctiondecompo})
gives an explicit connection between periodic ray trajectories in the
elliptical disc and the eigenfrequencies of the system. By taking the
Fourier transform of $\mathcal{N}_{\mathit{oscill}}(k)$, one should be able to
recover the periodic trajectories including orbits that change polarization
along their path. 
\begin{figure}[h]
\includegraphics[scale=0.7]{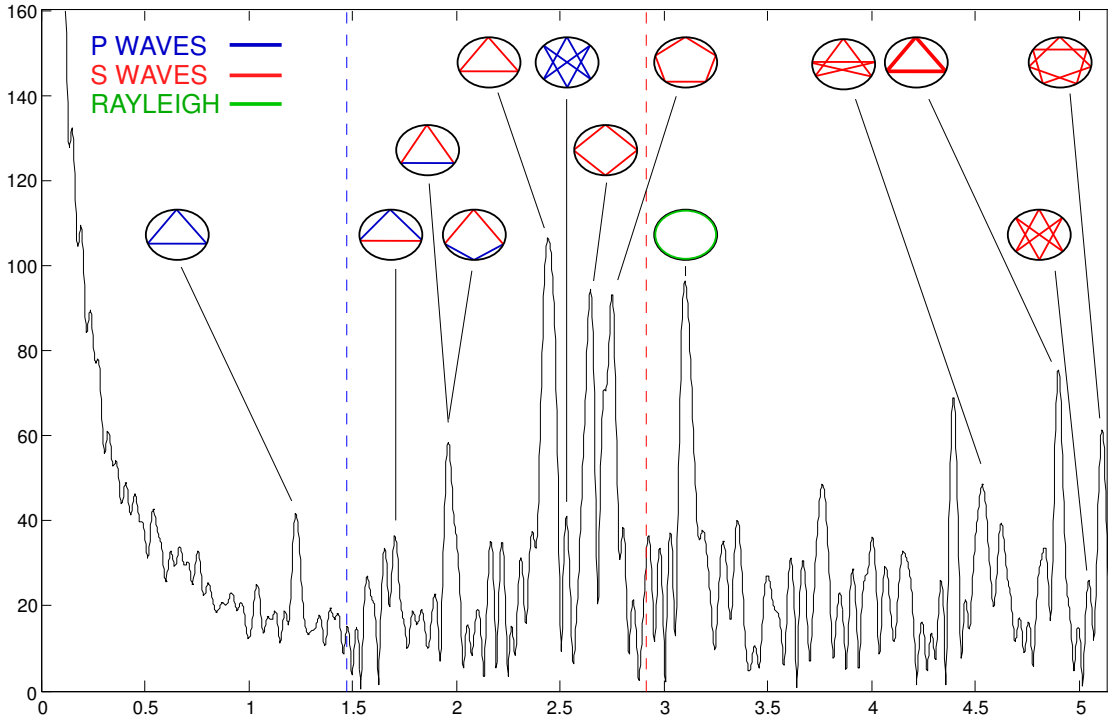}  \centering
\caption{Periodic orbits spectrum of an aluminium elliptical disc $e=0.6$ ($%
\frac{b}{a}=0.8$).}
\label{fig:op}
\end{figure}
Figure \ref{fig:op} shows a numerically calculated period spectrum obtained
from the first $1540$ resonances of an aluminium elliptical disc (Ag4mc - $%
c_{L}=6344.8$ m.s$^{-1}$, $c_{T}=3196$ m.s$^{-1}$) in the frequency range $%
0<k\leq 150$. We remove the smooth part of $\mathcal{N}(k)$ and take the
Fourier transform. The thus obtained period spectrum shows numerous \textit{%
peaks} that correspond to: orbits being of pure pressure, pure shear, mixed
polarization types, and pure Rayleigh orbit. The two dashed lines correspond
to accumulation towards a limit orbit when segments of pure pressure or pure
shear become tangential to the boundary. For example, orbits such as the
irregular pentagram or those that turn twice (appearing at $\approx 4.9$)
can also be resolved. We note that orbits with segments of both pressure and
shear polarization type arise when mode conversion occurs. Surface Rayleigh
orbit (appearing at $\approx 3.1$) can also be clearly identified including
in spectra computed for other various eccentricities. The \textit{lengths}
of periodic orbits of pure pressure and pure shear are obtained using the
expressions given by Sieber \cite{sieber}, whereas those undergoing mode
conversion are geometrically obtained using Snell-Descartes law. The
abscissa axis actually corresponds to dimensionless travel time due to the
chosen frequency normalization.

\section{Conclusion and perspectives}

From symmetry considerations, a classification of resonances of the elastic
elliptical disc has been provided: they lie in four distinct families
associated with the four irreducible representations $A_{1},A_{2},B_{1},B_{2}
$ of the symmetry group $\mathcal{C}_{2v}$ of the elliptical disc. Each
resonant mode is labelled by the two numbers $\left( n,\ell \right) $ in the
modal formalism associated to the circular disc. They are also tagged by the
associated irreducible representation. The splitting up of resonances is
emphasized and algebraic considerations permit to understand this
phenomenon. This method significantly simplifies the numerical treatment of
the problem. The computations can be also carried out in case of large
eccentricities when rapid variations of the curvature radius occur.

A connection between the resonances spectrum and the periodic orbits
spectrum is established. Due to mode conversion, orbits changing
polarization along their path appear. The pure Rayleigh orbit is also
clearly resolved for various eccentricities.

Finally, a series of experiments will be added to the theoretical studies.
An experimental part based on laser impacts excitation and laser vibrometry
will be carried out for three-dimensional objects. We expect to observe
splitting up of resonances and resonant modes crossings when the sphere is
deformed to the spheroid (symmetry breaking $\mathcal{O}(3)$ $\rightarrow $ $%
\mathcal{D}_{\infty h}$), as we theoretically observe in the two-dimensional
case when the circular disc is deformed to the elliptical one.

The method described in this paper is also well suited to scattering
problems such as acoustic scattering by infinite elastic cylinders of
elliptical cross-section (work in preparation). In this context, further
experiments in underwater acoustic scattering will also be performed.

\appendix

\section{Coefficients}

\label{coefficients}

The structural functions involving Bessel functions are angular dependent.
They are given by%
\begin{eqnarray}
c_{q}^{L}& =&(x_{T}^{2}-2q^{2})J_{q}(x_{L})+2x_{L}J_{q}^{\prime }(x_{L}) \\
d_{q}^{L}& =&2q(x_{L}J_{q}^{\prime }(x_{L})-J_{q}(x_{L})) \\
c_{q}^{T}& =&f_{q}^{T}=2q(x_{T}J_{q}^{\prime }(x_{T})-J_{q}(x_{T})) \\
d_{q}^{T}& =&(x_{T}^{2}-2q^{2})J_{q}(x_{T})+2x_{T}J_{q}^{\prime }(x_{T}) \\
f_{q}^{L}& =&(2x_{L}^{2}-x_{T}^{2}-2q^{2})J_{q}(x_{L})+2x_{L}J_{q}^{\prime
}(x_{L})
\end{eqnarray}%
where $x_{j}=k_{j}r(\theta )$, $j\in \{L,T\}$, $q\in \mathbb{N}$ (even or
odd).

\section{Fourier coefficients}

\label{fouriercoeff}

The Fourier coefficients are evaluated on the restricted fundamental domain,
they are numerically calculated from the following expressions%
\begin{eqnarray}
\beta _{q,r}^{j}& =&\frac{2}{\pi }\int_{0}^{\frac{\pi }{2}}[c_{q}^{j}\cos
(q\theta )\;n_{\rho }+d_{q}^{j}\sin (q\theta )\;n_{\theta }]\cos (r\theta
)d\theta \\
\xi _{q,r}^{j}& =&-\frac{2i}{\pi }\int_{0}^{\frac{\pi }{2}}[d_{q}^{j}\sin
(q\theta )\;n_{\rho }-f_{q}^{j}\cos (q\theta )\;n_{\theta }]\sin (r\theta
)d\theta \\
\alpha _{q,r}^{L}& =&-\frac{2i}{\pi }\int_{0}^{\frac{\pi }{2}}[c_{q}^{L}\sin
(q\theta )\;n_{\rho }-d_{q}^{L}\cos (q\theta )\;n_{\theta }]\sin (r\theta
)d\theta \\
\alpha _{q,r}^{T}& =&-\frac{2i}{\pi }\int_{0}^{\frac{\pi }{2}%
}[-c_{q}^{T}\sin (q\theta )\;n_{\rho }+d_{q}^{T}\cos (q\theta )\;n_{\theta
}]\sin (r\theta )d\theta \\
\eta _{q,r}^{L}& =&\frac{2}{\pi }\int_{0}^{\frac{\pi }{2}}[-d_{2n}^{L}\cos
(q\theta )\;n_{\rho }-f_{q}^{L}\sin (q\theta )\;n_{\theta }]\cos (r\theta
)d\theta \\
\eta _{q,r}^{T}& =&\frac{2}{\pi }\int_{0}^{\frac{\pi }{2}}[d_{q}^{T}\cos
(q\theta )\;n_{\rho }+f_{q}^{T}\sin (q\theta )\;n_{\theta }]\cos (r\theta
)d\theta
\end{eqnarray}
$j\in \{L,T\}$, $\left( q,r\right) \in \mathbb{N}^{2}$ (even or odd).

\end{document}